# High Pressure and Temperature Neural Network Reactive Force Field for Energetic Materials

Brenden W. Hamilton[1], Pilsun Yoo[2], Michael N. Sakano[3], Md Mahbubul Islam[4], Alejandro Strachan[5] *

[1]Theoretical Division, Los Alamos National Laboratory, Los Alamos, New Mexico 87545, USA
2 Computational Science and Engineering Division, Oak Ridge National Laboratory, 1 Bethel Valley Road, Oak Ridge, TN, 37830, USA
[3] Sandia National Laboratories, Albuquerque, New Mexico 87123 USA
4 Department of Mechanical Engineering, Wayne State University, Detroit, Michigan 48202 USA
5 - School of Materials Engineering and Birck Nanotechnology Center, Purdue University, West Lafayette, Indiana, 47907 USA

## Abstract

Reactive force fields for molecular dynamics have enabled a wide range of studies in numerous material classes. These force fields are computationally inexpensive as compared to electronic structure calculations and allow for simulations of millions of atoms. However, the accuracy of traditional force fields is limited by their functional forms, preventing continual refinement and improvement. Therefore, we develop a neural network based reactive interatomic potential for the prediction of the mechanical, thermal, and chemical response of energetic materials at extreme conditions for energetic materials. The training set is expanded in an automatic iterative approach and consists of various CHNO materials and their reactions under ambient and under shock loading conditions. This new potential shows improved accuracy over the current state of the art force fields for a wide range of properties such as detonation performance, decomposition product formation, and vibrational spectra under ambient and shock loading conditions.

* Corresponding author: strachan@purdue.edu

# 1. Introduction

Utilizing fundamental physics and chemistry to make predictions of the safety and performance of energetic materials remains critical scientific challenge that will enable faster development of materials and better engineered systems that employ such materials. Performance metrics, such as detonation velocity and pressure, are relatively easy to estimate as they are, to a large extent, determined by the equations of state of the reactants and products[1–4]. These quantities can be predicted using a range of techniques, such as molecular dynamic (MD)[1,5–7], thermochemical calculations[2,8,9], and machine learning models[3,4,10,11]. Crystal phases/structures and molecular conformation predictions for high explosive (HE) materials typically require higher complexity models but are computationally feasible[12–15]. Predictions of material safety and sensitivity are the most challenging, as they are often dependent on material microstructure and require the characterization of processes with a wide range of scales occurring at extreme conditions[16–18].

In the last two decades, MD studies have advanced our understanding of the shock compression and initiation of chemical reactions in high explosives[15,19–29]. MD-informed multiscale modeling efforts have recently enabled significant advances in studying shock compression and deformation in HEs, especially focusing on the thermo-mechanics of hotspot formation[26,30–32]. Nonreactive MD simulations in HE crystals have unraveled numerous deformation mechanisms and routes for shock-induced energy localization[25,33–40]. While these force field-based simulations provide an accurate description of thermo-mechanical properties and performance, uncertainties remain regarding their description of detailed chemistry, especially under shock loading[19]. Ab initio and quantum chemistry methods provide a more accurate description of atomic interactions but at a steep computational cost. Electronic structure simulations allowed for predictions of reaction pathways and kinetics of explosives from both thermal and shock loading[15,29,41–44]. However, the length and timescales of density functional theory (DFT) and density functional based tight binding (DFTB) MD have limited studies to small systems and gas phase reactions, preventing the influence of microstructure and crystalline defects, forcing a reliance on classical potentials for these studies that are crucial to a materials level understanding of energetic materials.

The development of reactive interatomic potentials such as ReaxFF[45,46], whose computational cost scales linearly with the number of atoms, has allowed for reactive MD simulations on the orders of thousands to tens of millions of atoms with current computational capabilities. From the first reactive shock simulations with ReaxFF[22], the predictions of reaction kinetics and pathways have been performed for numerous HE materials[6,7,23,47,48], and significantly more complex studies of initiation and reaction have been enabled from the potential's development. ReaxFF simulations of reactive pore collapse have enabled the direct study of hotspot formations and transition to deflagration[24,26,49,50]. ReaxFF has also enabled direct comparisons to *in situ* infrared (IR) spectroscopy experiments in shocked HEs, helping to validate the accuracy of the potential[51–53]. Recently, more complex studies have been performed, exploring the molecular scale influences on directional sensitivity[54], mechanochemistry in hotspots[55,56], and upscaling MD chemistry to the continuum scale[57].

The original ReaxFF parametrization was designed for hydrocarbons and CHNO systems, but its parametrization has been extended to a large portion of the periodic table[46] and applications range from biology[58] to electrochemistry[59]. The calculation of covalent interactions in ReaxFF is based on partial bond orders (BOs), where the bond order between two atoms is a many-body function of local chemical structure[45]. Bond-stretch, angle, and torsion terms depend on bond-orders and over- and under-coordination penalties control the number of bonds that can be formed. The total energy is assumed to be a sum of various terms:

$$E_{system} = E_{bond} + E_{over} + E_{angle} + E_{tors} + E_{vdWaals} + E_{Coul} + E_{Specific}$$

where $E_{Specific}$ are additional terms specific materials. One example of this is the low gradient term used to correct for long-range London dispersion not captured via the original Morse function[60]. This leads to significantly more accurate densities for molecular crystals such as HEs, lowering the average equilibrium volume error by an order of magnitude. This parametrization is denoted ReaxFF-LG. Additional modifications to predict better densities and equations of state was to provide further inner-core repulsions for the van der Waals interactions,[61] this is denoted ReaxFF-IW. Wood et. al. developed a CHNO parametrization for nitramines (a class of energetic materials) based on the training sets for the original CHNO parametrization and a combustion training set, which included full disassociations of 1,3,5,7-Tetranitro-1,3,5,7-tetrazocane (HMX), 1,3,5-Trinitro-1,3,5-triazinane (RDX), and pentaerythritol tetranitrate (PETN)[62]. We denote this parametrization as ReaxFF-2014. ReaxFF-2014 was further appended as the energetics and combustion branch merger from Ref [24] and added an LG correction, which we denote as ReaxFF-2018[26]. While significant scientific advances have been made with ReaxFF, and parametrizations have improved the overall description of CHNO systems, it is limited by its physics-inspired functional form and well-documented deficiencies have persisted[52].

Recent interest in using machine learning (ML) methods for science resulted in the development of ML interatomic potentials[63,64]. In this approach, physics-agnostic models relate the local atomic environment surrounding every atom, described with fingerprints that embed physics, to the energies and forces[65–68]. In this work we expand upon and demonstrate the use a neural network reactive force field (NNRF) for CHNO systems[69]. NNRF uses atom-centered, weighted-Gaussian, symmetry functions developed by Gastegger et al.[70] and a high-dimensional neural network potential (HDNNP) suggested by Behler and Parrinello[71].

The original parametrization of NNRF for CHNO utilized an iterative approach to enhance the training set of atomic configurations used to train the NN. Forces and energies for training and testing were recalculated using DFT. The initial training set included snapshots for a decomposition simulation of RDX using ReaxFF, bond dissociation for all di-atomic molecules in the CHNO set, and various states from simulations of isolated molecules chosen from known RDX intermediates and products. The training set was then continuously expanded via an active learning loop in which states explored by MD simulations using latest NNRF were added to the training set with their DFT energies and forces. This allowed for the NNRF to explore and find bad energy states that it poorly represented, allowing them to be retrained to correct them.

The final iteration in Ref. [69], Gen1.9, showed accurate results for thermochemistry and structural information, as well as easy transferability to other HEs by expanding the training set. However, NNRF Generation 1 was not parametrized to handle high-pressure chemistry, which is vital to understanding HEs under some of their typical operating conditions such as shock initiation and detonation. This paper describes a significant expansion of the training set and applicability of NNRF. NNRF Gen 2.X was trained with the addition of thermal decomposition and 300 K equations of state for HMX, nitromethane (NM), PETN, 2-Methyl-1,3,5-trinitrobenzene (TNT), 2,4,6-Trinitrobenzene-1,3,5-triamine (TATB), and Hexanitrohexaazaisowurtzitane (CL20). For Gen3.X we added thermal- and shock-induced decomposition of PETN between 0 GPa and 70 GPa, where Gen3.9 included a ZBL potential to correct predictions of short bond distances. We use this new Generation 3 NNRF to predict the detonation velocity and CJ pressure of numerous HEs, as well as their reaction product formations and IR spectra, all which show equal or superior accuracy to ReaxFF, despite only being under development for 1/10$^{th}$ of the time.

## 2. Methods

### NNRF training

Training of the NNRF forcefield involves an iterative loop that generates progressively more relevant configurations using an automated procedure that can be easily generalized to other systems. The first generation[69] started with isothermal-isochoric thermal decomposition reactions of RDX using ReaxFF-2014. Snapshots were extracted every 0.1 picosecond (of a 400 picoseconds run) and their forces re-calculated using DFT with a Perdew–Burke–Ernzerhof (PBE) exchange-correlation functional[72] and the empirical D2 correction for London dispersion interactions[73]. These positions and forces, along with bond dissociation energies for CHNO systems and isolated molecules of key RDX products and intermediates make up the initial training set. Each iteration within each generation adds additional trajectory frames to the training set in an iterative loop using simulations of HE decomposition (or shock in later generations) with the NNRF forcefield from the previous iteration. DFT forces are then calculated on trajectory frames and unique molecules are extracted for isolated calculations, adding both into the training set. This iterative learning approach is repeated until testing error reach a reasonable low level.

In this paper, the Gen 1.X training sets were expanded to include, for Gen2.X, thermal decomposition and 300 K equation of state for HMX, NM, PETN, TNT, TATB, and CL20. Gen3.X adds high pressure and shock induced decomposition of PETN between 0 GPa and 70 GPa. Figure 1 shows the number of structures and their characteristics for each generation. We further add to and correct high density behavior using the ZBL potential to form Gen3.9zbl, as the direct training of high pressure and density states without a ZBL potential led to overly high energies at short bond distances that was meaningful error at experimentally relevant pressure states. The NNRF potential and datasets for the Gen1.9, Gen2.8 and Gen3.9zbl potentials are available at the following github repository: https://github.itap.purdue.edu/StrachanGroup/nnrf_nitramines. Neural network architecture was identical to Gen1.9: 42 input nodes, 2 hidden layers of 50 nodes and 1 output for atomic energy prediction. For the input nodes the atomic structures were converted to local environment descriptors using weighted gaussian symmetry functions[70].

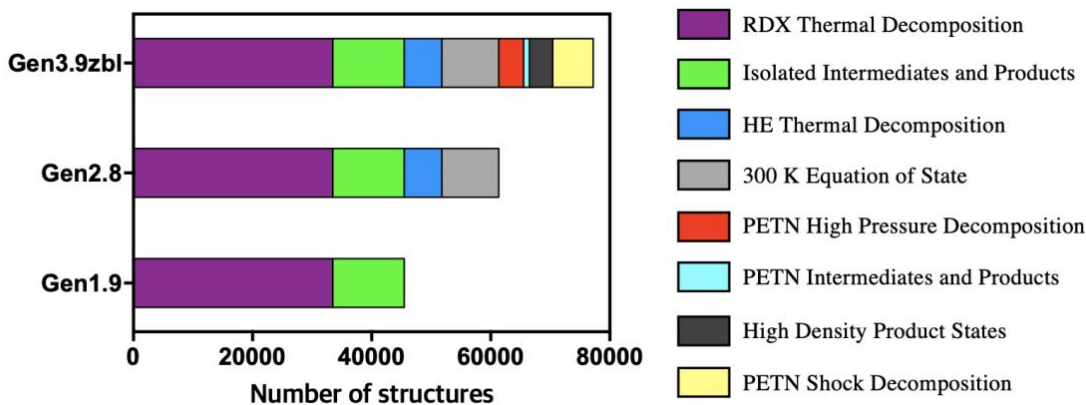

*Figure 1: Distribution of number of structures in the training sets from various MD simulation conditions and molecules included.*

## Molecular dynamics simulation tests of NNRF

For MD comparisons of NNRF to ReaxFF, five interatomic potentials are used here: NNRF Gen 3.9zbl (this work), ReaxFF-2018[26], ReaxFF-2014[62], ReaxFF-LG[60], and ReaxFF-IW[61]. Eight high explosive (HE) materials were used: RDX, HMX, TNT, TATB, CL-20, NM, PETN, and polyvinyl nitrate (PVN). All MD simulations are conducted using a 0.1 femtosecond timestep and charge equilibration with QEq[74] with a accuracy tolerance of $1 \times 10^{-6}$. Thermal decomposition simulations are conducted at the potential's relaxed density with a Nose-Hoover thermostat[75] in which the relaxed density is reached using an anisotropic barostat (or triaxial in the case of non-orthorhombic cells). Shock simulations were conducted with the constant stress Hugoniostat[76]. Initial product states to map the product Hugoniot were created by shocking the system to 50 GPa and running dynamics until the reaction reached full exothermicity (temporally steady values of PE and KE). This typically took on the order of hundreds of picoseconds, dependent on the HE. A pressure of 30 GPa was used for nitromethane and polyvinyl nitrate due to numerical instabilities in the product states above 40 GPa.

To map the product Hugoniot curves, additional (release wave) Hugoniostat simulations were conducted on the products. These product Hugoniots were fit to a power law function $P(X) = A * X^B$ and the Rayleigh line for steady state detonation velocity was given the form $P(X) = a * x + b$ where $X = \frac{V}{V_o} = \frac{\rho_o}{\rho}$. This results in the solution states, from the fit Hugoniot:

$$X_{CJ} = \frac{B}{B-1}$$

$$a = A * B * X_{CJ}^{B-1}$$

$$b = -a$$

$$P_{CJ} = P(X_{CJ}) = a * X_{CJ} - a$$

$$D_V = \sqrt{\frac{P_{CJ}}{\rho_o * (1 - X_{CJ})}}$$

IR spectra are calculated from Hugoniostat simulation trajectories using the same method presented in Ref. [53]. The Generalized Crystal-Cutting Method[77] was used to align the original PETN [110] crystallographic direction along the cartesian Z axis which was the shock direction. The IR spectrum was generated using the time evolution of the charge moments[78]:

$$I(\omega; t_0) = \frac{2\pi\omega}{3\hbar c n}(1 - e^{-\hbar\omega/k_B T}) \sum_{n=-F/2}^{F/2} e^{-i2\pi\omega(t_0+n\Delta t)} \left[\frac{1}{M-j}\sum_{i=1}^{M-j} \dot{M}(i) \cdot \dot{M}(i+j)\right]$$

$$\dot{M}(i) = \sum_{j=1}^{N} q_j(i) \cdot v_j(i)$$

where $\omega$ is the wavenumber, $t_0$ is the timeframe around which the calculation is centered on, $\hbar$ is Planck's constant divided by $2\pi$, $c$ is the speed of light, $N$ is the number of atoms in the system, $k_B$ is the Boltzmann constant, $T$ is the system temperature, $F$ is the number of times sampled in the analysis (2048 frames), $\Delta t$ is the sampling rate, and $q_j(i)$ and $v_j(i)$ are the charge and velocity of atom $j$ at timeframe $i$, respectively. For this particular application, charges on the hydrogen and carbon atom were set to 0, such that the IR spectra depended only on nitrogen and oxygen atom types. The time-evolved IR spectra was computed at 1 picosecond intervals using a running average composed of 5 picoseconds of simulation time each. Due to the Hugoniotstat technique enforcing a uniform shock on the entire system, a time convolution using a weighted sum between the shocked and unshocked IR spectra was implemented. The convolution method takes into account the sample thickness and shock velocity to determine to account for the time delay required to fully shock the material as observed in the experiments, for further details, see Ref. [52].

## 3. NNRF accuracy over the training set and the QM9 database

Figure 2 shows parity plots for energies and forces comparing selected generations of NNRF with DFT-PBE-D2, with RMS error values inset in the figure. The test set in this plot consists of isothermal-isochoric decomposition reactions and shock-induced (via Hugoniostat simulations) reactions for RDX, HMX, CL-20, PETN, TNT, and TATB, at the densities and pressures shown in Figure 3 which displays individual cases. Significantly larger error reductions were made from Gen1.9 to Gen 2.8 than from Gen 2.8 on. This is due to the expansion of the training set to other HE formulations and the 300 K EOS for these solids. Since Gen1 was only trained on ambient density, it was not expected to perform well under shock and high-pressure conditions. The improvements from Gen2.8 to Gen 3.9zbl can be attributed to both training on reactions in higher density states and the inclusion of the zbl reference potential which gives considerably better predictions at small separation distances.

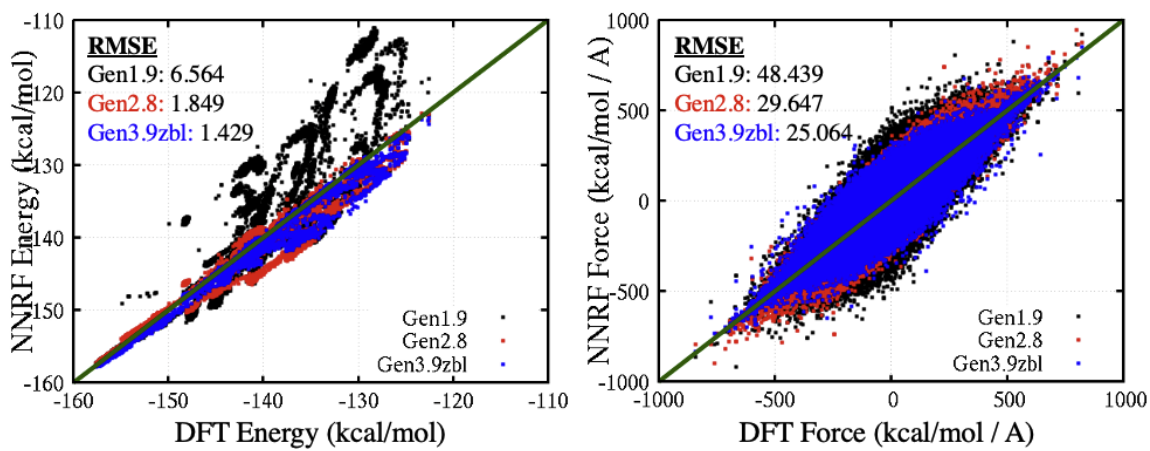

*Figure 2: Energy and Force testing set parity plots for NNRF simulations including high-pressure states.*

Figure 3 highlights specific cases of high-pressure thermal decomposition and shock initiation for HMX and CL-20, with both cases significantly lowering the RMSE, often by 50% in the highest-pressure cases from Gen1.9. The HMX ambient density case performs well for Gen1.9 due to the similarities of HMX to RDX which the potential was based on. On the other hand, CL-20, a quite disparate molecule, is not as easily extrapolatable from a nitramine-based training set. While the initial Gen1.9 was not perfectly transferable to all HE materials, it does work well at ambient conditions for nitramines. However, for both molecules, increased pressure causes significant errors. Progressing onto Gen3.9zbl, both high-pressure states and a wider range of energetic molecules perform better.

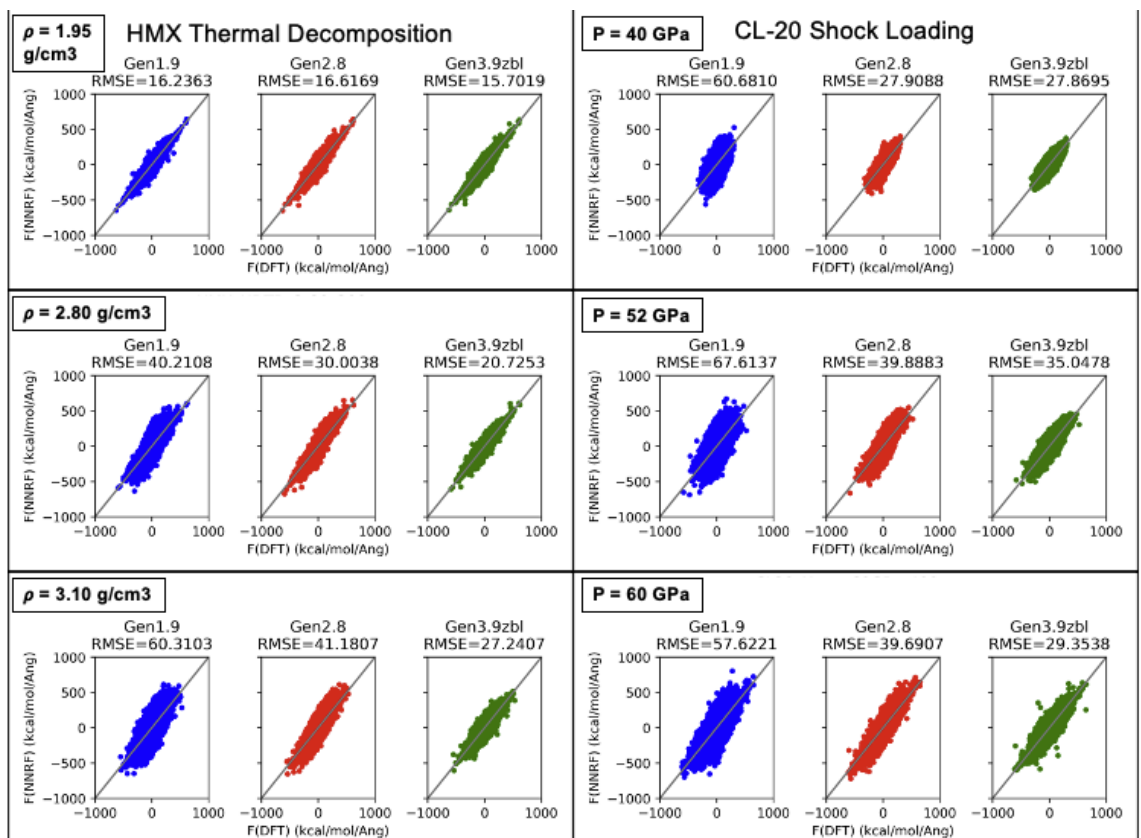
*Figure 3: Force parity plots for all three generations for various high pressure decomposition and shock initiation simulations.*

Table 1 shows RMS error in predictions of formation energy for PBE-D2, all three NNRF generations, and all four ReaxFF parametrizations for the eight HEs used here, compared against the experimental values. Interestingly, NNRF Gen2.8 has the worst values, with Gen3.9zbl improving on both previous NNRF versions. NNRF Gen3.9zbl also does better than all four ReaxFFs, but does not achieve the PBE-D2 level accuracy.

*Table 1: RMS Errors in the prediction of formation energies for the 8 HEs used in this work from experiments.*

| Model | Formation Energy RMSE (kcal/mol) |
|---|---|
| PBE-D2 | 28.11 |
| NNRF Gen1.9 | 79.82 |
| NNRF Gen2.8 | 267.98 |
| **NNRF Gen3.9zbl (this work)** | 42.71 |
| ReaxFF-2018 | 60.87 |
| ReaxFF-2014 | 48.58 |
| ReaxFF-LG | 56.54 |
| ReaxFF-IW | 54.92 |

Figure 4 shows parity plots for formation energies of the 133,885 molecules included in the QM9 dataset[79] for NNRF Gen3.9zbl and the four ReaxFF parametrizations. The formation energies predicted by the force fields are compared with the B3LYP/6-31G(2df,p) results included in QM9. Each panel also contains its RMS error in kcal/mol/atom as inset text, NNRF Gen3.9zbl represents a considerable improvement over the ReaxFF predictions. This shows the potential of an improved transferability of NNRF, as compared to these versions of ReaxFF, to non-HE systems and processes.

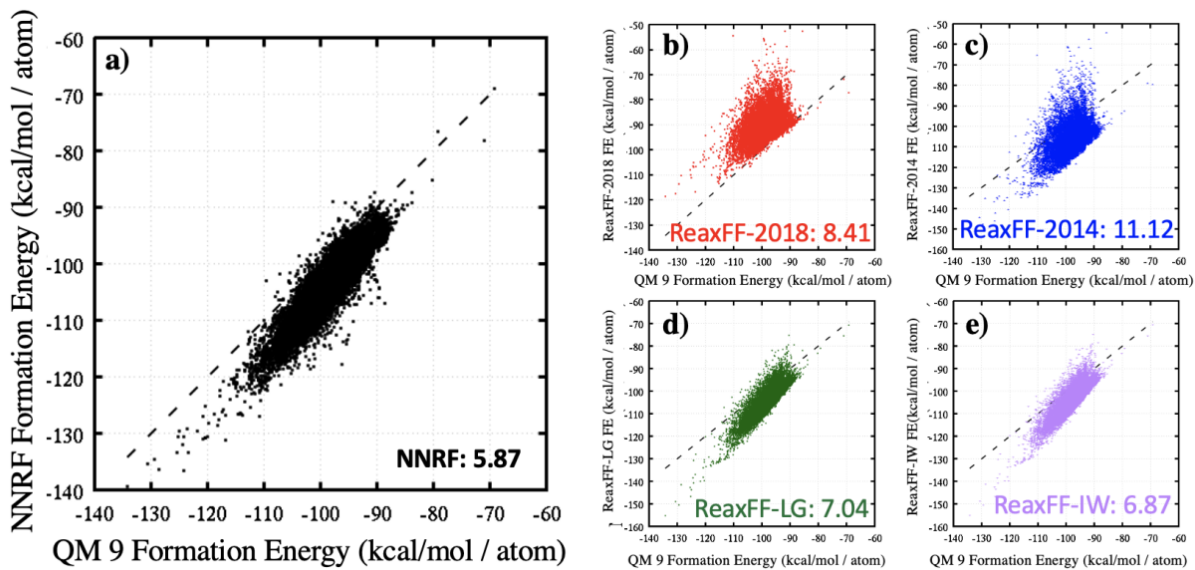

*Figure 4: Parity plots of formation energies for NNRF Gen3.9zbl and the four ReaxFF parametrizations used here for the QM9 dataset. Inset text is RMSE values in units of kcal/mol/atom.*

## Prediction of Performance Properties of HE materials

Using the Hugoniostat method, one can calculate the reagent and products Hugoniot curves, as described in Methods, and obtain detonation velocity and pressure. The reagent Hugoniot was obtained from independent runs at different shock pressures starting from ambient conditions. The product Hugoniot from subsequent Hugoniostat runs as additional compression or release waves on a product gas state from a strong initial shock run. Figure 5 shows selected reagent curves in the $U_s$-$U_p$ plane. The HEs shown were chosen due to the availability of experimental results that span the relevant shock strengths studied here and for non-formulation systems (single crystal or neat powders). The NNRF data (points) compares well to the experimental values for slope and intercept which are related to the materials' sound speed and pressure derivative of the bulk modulus, as well as elastic constants/orientation in anisotropic materials like TATB and PETN.

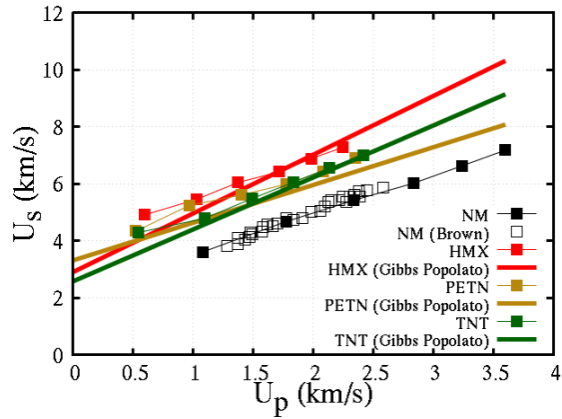

*Figure 5: Reagent Us-Up plots from 3.9zbl NNRF simulations compared to experiments from Refs [80,81].*

More relevant to the performance of an energetic material is the accuracy of the product Hugoniot, which involves obtaining the product gas species correctly as well as its thermo-mechanical properties. From fitting these curves in P-V space to a power law function and solving for a tangential Rayleigh line that also intersects the initial P-V state of the unshocked material, CJ Pressure and detonation velocity can be calculated[1]. These are the tangential point and related to the slope of the Rayleigh line, respectively. Figure 6 shows the predicted detonation velocity and CJ pressure for all eight HEs using NNRF Gen3.9zbl and the four ReaxFF parametrizations. RMS errors as compared to experimental values are included as inset text. For detonation velocity, NNRF improves on three of the four ReaxFF parametrizations. ReaxFF-2018, which makes the better prediction of detonation velocity, also has slightly better predictions of ambient density, which affects the x-intercept of the Rayleigh line used. Interestingly, however, the poorest detonation velocity prediction from NNRF, with respect to ReaxFF-2018, is the liquid nitromethane. For CJ pressure, NNRF Gen3.9zbl outperforms all ReaxFFs, but only minor improvements over ReaxFF-2018. The tangential point, which is used for CJ pressure, is less sensitive to the initial density state in general, making the CJ pressure predictions more dependent on the accuracy of the final product state, the reaction paths and the thermodynamics predicted by each model. In the power law model described in Methods, $P(X) = A * X^B$, the value of A is significantly more sensitive to $\rho_o$ changes than B. From the HMX Hugoniot, ranging the initial density from 1.70 to 2.00 g/cm$^3$, the relative CJ pressure ranges from 0.611 to 1.573, but relative detonation velocity only ranges from 0.816 to 1.206, plots are shown in Supplemental Material section SM-1. This more greatly weighs the error in CJ pressure to be from error in the gas product mixture as compared to the detonation velocity error which will be heavily affected by $\rho_o$ error.

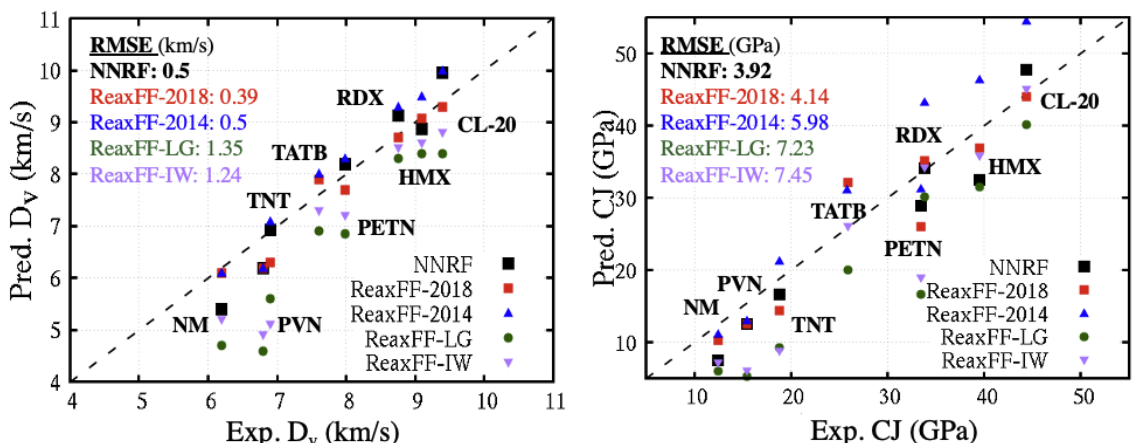

*Figure 6: Detonation velocity and CJ pressure parity plots for NNRF and all four ReaxFF parametrizations used here, calculated from the product Hugoniot curves. Experimental values taken from Refs. [82].*

Figure 7 compares the relative amounts of critical product molecules obtained via isothermal decomposition of multiple HE materials. The decomposition simulations are conducted at 2500 K and ambient density. We show results for NNRF Gen3.9zbl, the various ReaxFF parametrizations, and experiments. The experimental results were obtained from Ref. [82], in which the experiments are typically not in the isochoric condition as done with MD, therefore expecting lower yields across the board from MD. Thus, comparison against experiments should be done with care, with a focus on trends rather than exact one-to-one numerical values. In general, ReaxFF-LG and ReaxFF-IW do better than ReaxFF-2018 and ReaxFF-2014 when compared to experimental values. NNRF Gen3.9zbl is, at least, on the order of ReaxFF-2018 and ReaxFF-2014, or better, for $N_2$ and $H_2O$ production, is moderate to good compared to all four ReaxFFs for $CO_2$ production, and consistently underpredicts $NH_3$ production.

Figure 8 shows the final NNRF product amounts for each HE from isothermal decomposition, per initial molecule, also scaled by the molecular weight of the HE. The non-scaled plot is provided in Supplemental Material section SM-2. TATB, which forms significant carbon condensates and large heterocycles than can trap N and O[28,83], shows the lowest total gas product formation. Higher performance HEs like CL-20, PETN, and HMX show significantly more normalized product yields, showing the NNRF results at least track with basic chemical intuition.

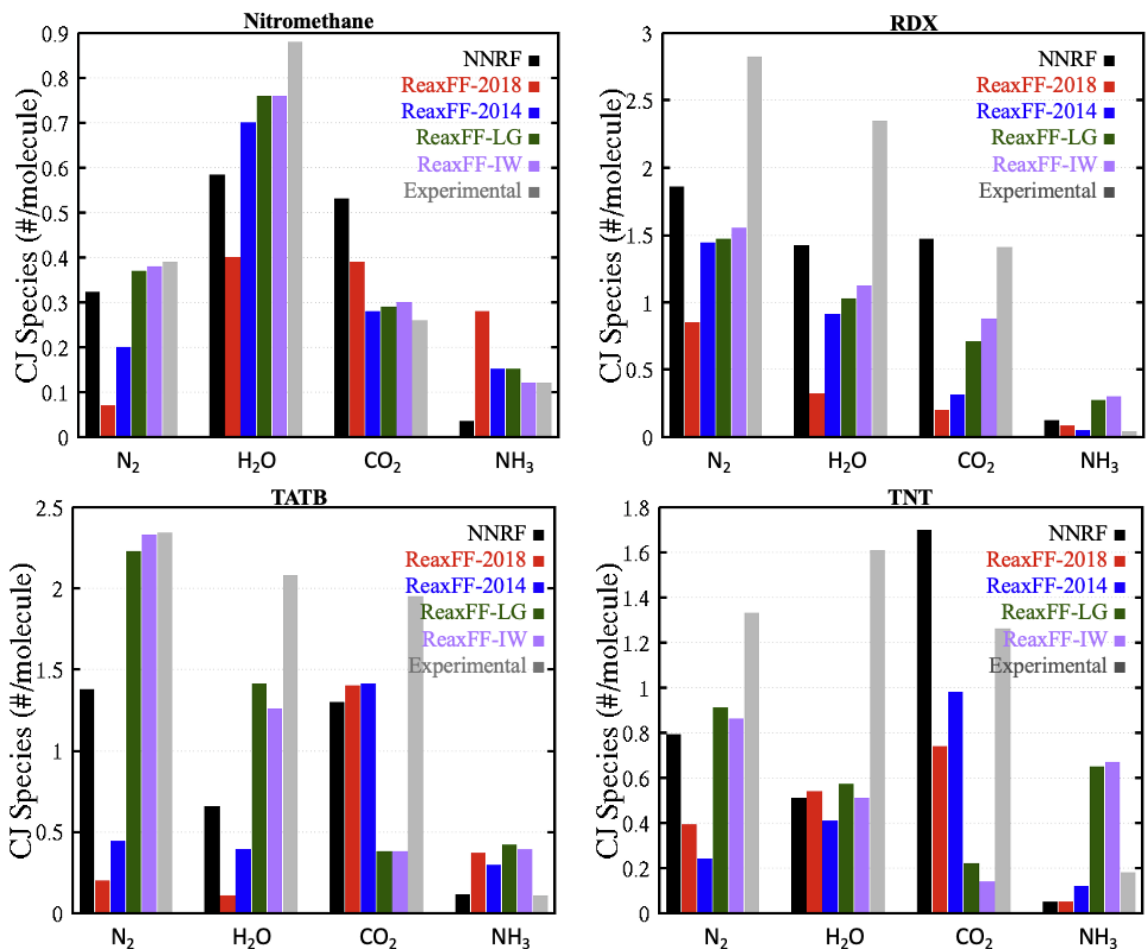

*Figure 7: NNRF and ReaxFF final products for isothermal decomposition simulations at 2500 K, compared to relevant experiments, for key product species. Experimental values are from Ref [84].*

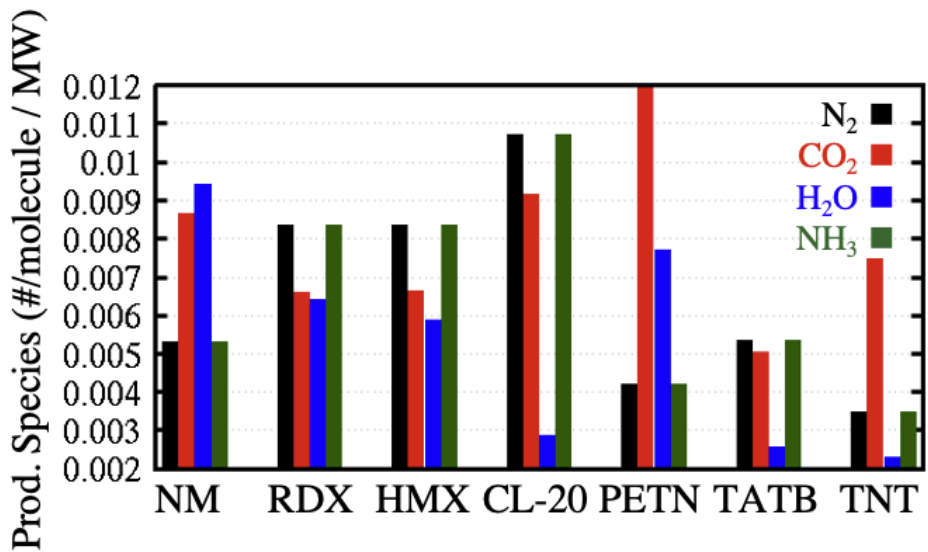

*Figure 8: NNRFGen3.9zbl final thermal decomposition product amounts per HE molecule, scaled by the molecule's molecular weight.*

## Vibrational spectra and shock-induced chemical reactions

The roto-vibrational density of states (DoS) describes the frequencies associated with normal models or phonons of a material and is an important material property that governs the thermal contribution to the free energy and the transfer of energy between modes during thermal equilibration. Infrared (IR) active modes can be measured via spectroscopy, and ultrafast IR spectroscopy has been used to experimentally assess shock-induced chemistry[53,81,85,86]. While challenging, these experiments provide the most direct measure of detailed chemistry under shock loading. The analysis of MD simulations can provide the same observable and can help interpret experiments. Comparisons have shown that ReaxFF can describe the timescales and threshold shock strength for the onset of chemistry quite accurately but the details of the evolution of the IR spectra during shock loading in the simulations differed significantly from experiments[19].

Figure 9 shows the DoS for four materials obtained from the autocorrelation of atomic velocities from Gen3.9zbl at ambient conditions. Of interest is the symmetric $NO_2$ stretch ($v_s$-$NO_2$) and anti-symmetric $NO_2$ stretch ($v_a$-$NO_2$) modes, as this specific bond is thought to play an important role in the initial decomposition of these energetic molecules. For PETN, Gruzdkov and Gupta compared simulated Raman spectra against their own experiments[87]. Gen3.9zbl shares common overlapping frequencies for $v_a$-$NO_2$ around ~1700cm$^{-1}$ and $v_s$-$NO_2$ stretches around ~1300cm$^{-1}$, which is an improvement over DoS using ReaxFF constructed by Wood and Strachan[88]. For nitromethane, Liu et al. calculated vibrational spectra in the solid phase using DFT. In particular, both $v_s$-$NO_2$ and $v_a$-$NO_2$ stretch modes at ~1400cm$^{-1}$ and ~1550cm$^{-1}$, respectively, show improvements to the calculated frequencies using ReaxFF[88]. Alzate et al. compared theoretical spectra of TNT using DFT with experiments, and noted $v_s$-$NO_2$ at ~1350cm$^{-1}$ and $v_a$-$NO_2$ at ~1550cm$^{-1}$; these frequencies match closely with NNRF3.9zbl[89]. Finally, PVN IR spectra observed by McGrane, Moore, and Funk saw $v_s$-$NO_2$ at 1270cm$^{-1}$ and $v_a$-$NO_2$ at 1625cm$^{-1}$, which is also observed using Gen3.9zbl[90]. On the other hand, ReaxFF greatly overpredicts both frequencies[51]. In summary, NNRF Gen3.9zbl compares well with experiments in terms of vibrational properties that, in turn, govern important materials properties. This is a significant improvement over the state of the art.

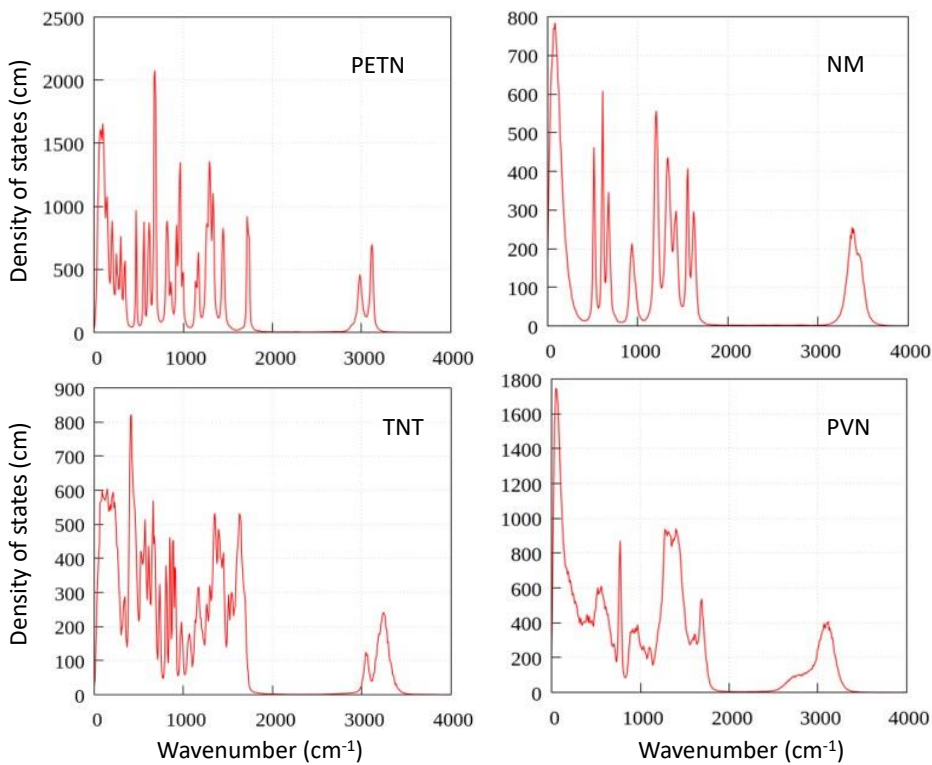

*Figure 9. Computed DoS using Gen3.9zbl for PETN, nitromethane, TNT, and poly(vinyl nitrate) at ambient density. Both symmetric and anti-symmetric $NO_2$ stretch modes show improvements to match experimental and DFT calculated frequencies. In contrast, ReaxFF can be off by as much as 300cm$^{-1}$.*

The evolution of the IR spectra with time for selected materials and shock strengths are shown in Figure 10. These were chosen due to the availability of experimental results for comparison. In the case of PVN, we observe the disappearance of the $NO_2$ peak between 100 and 150 picoseconds, in good agreement with experiments[85]. ReaxFF predicted similar behavior but overestimated the frequency of the mode[51]. In the case of PETN, NNRF predicts an increase in peak intensity and broadening of $v_s$-$NO_2$ and $v_a$-$NO_2$ stretch modes at around 150 picosecond, in good agreement with experiments[53]. On the other hand, ReaxFF-2014 shows a decrease in the antisymmetric stretch mode disappearing around 150 picosecond[53]. For TNT, NNRF predicts the weakening of the $NO_2$ peaks, while experimentally they seem to increase in intensity and broadening[91]. Finally, PVN spectra predicts a slight shift in the predicted $v_a$-$NO_2$ stretch mode (1680cm$^{-1}$), as compared to expectations from experiments[90] (1625cm$^{-1}$), but this is still an improvement over ReaxFF calculations[51] (1930cm$^{-1}$). In general, in the extremely challenging test of predicting the temporal evolution of IR spectra under shock conditions NNRF Gen.39zbl does not perfectly agree with experiments but shows marked improvements against state-of-the-art legacy reactive force fields.

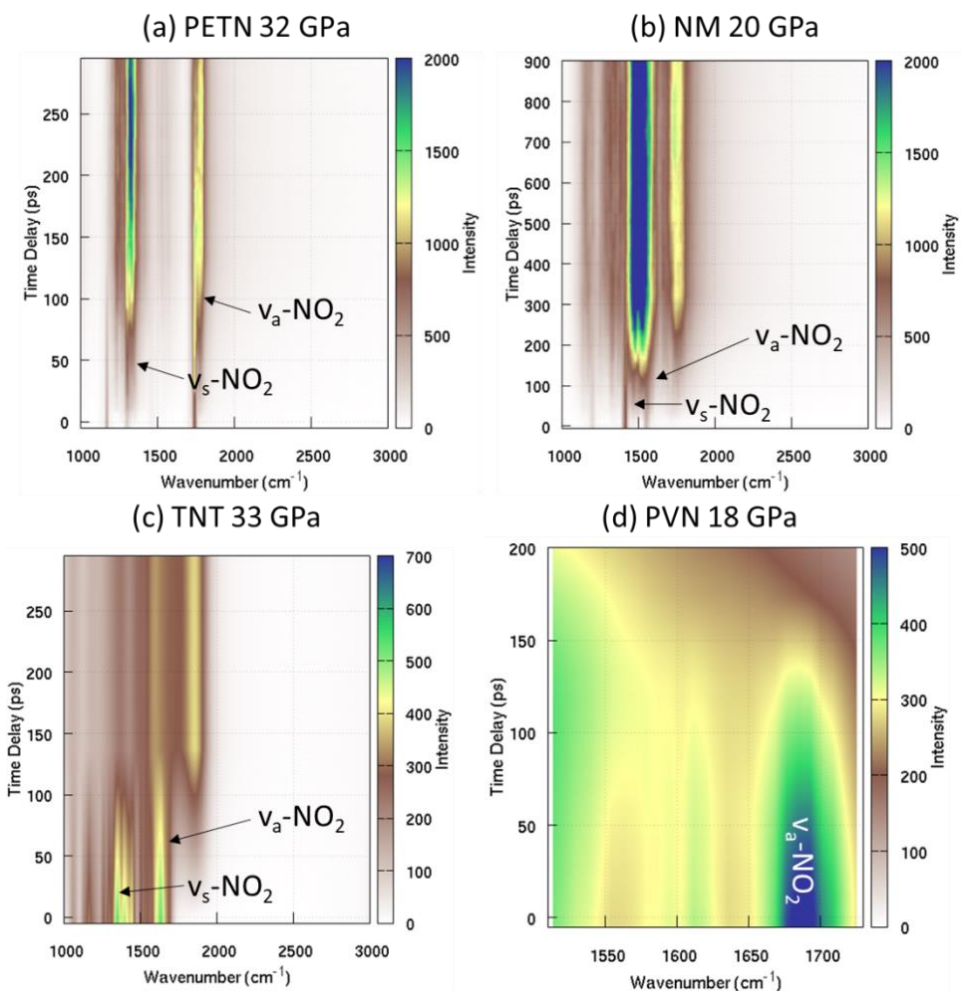

*Figure 10: Convoluted IR spectra for three energetic materials (a) PETN, (b) nitromethane, (c) TNT, and (d) poly(vinyl nitrate) under shock loading. In particular, symmetric and anti-symmetric $NO_2$ stretch modes are highlighted as they are thought to be the specific bond necessary to initiate chemistry.*

## Summary

Here, we developed a new generation of the Neural Network Reactive Forcefield (NNRF) that significantly improves the description of high-pressure and shock-loading properties over the state of the art. The initial Gen1 trained on only RDX decomposition reactions at ambient density. Gen2 adds ambient data and equations of state for a variety of energetic materials, and Gen3 added training data of shock-induced reactions in only PETN. NNRF Gen3.9zbl describes detonation properties and product concentrations in good agreement with experiments, with an accuracy matching the current state-of-the-art. However, Gen3.9zbl performs better than prior ReaxFF parametrizations in the prediction of formation energies not only of HE materials but over the entire QM9 dataset, showcasing NNRF's added flexibility ability to be utilized for a wider range of applications. Improved predictions of the vibrational density of states for various HE leads to a more accurate description of free energies and thermal transport. In addition, an improved description of the evolution of spectra under shock conditions indicates the possibility of developing definite models of chemistry under extreme conditions by contrasting experiments and reactive MD simulations.

Being based on neural networks, NNRF can be improved with additional data as long as the descriptors of local atomic environments can capture the different configurations encountered. Future work with other methods, including graph neural networks,[92] could result in additional improvements.

## Acknowledgements

This work was supported by the US Office of Naval Research, Multidisciplinary University Research Initiatives (MURI) Program, Contract: N00014-16-1-2557. Program managers: Chad Stoltz and Kenny Lipkowitz. Approved for Unlimited Release: LA-UR-23-21289.

# Supplemental Materials to: High Pressure and Temperature Neural Network Reactive Forcefield for Energetic Materials

Brenden W. Hamilton[1], Pilsun Yoo[2], Michael N. Sakano[3], Md Mahbubul Islam[4], Alejandro Strachan[5]

[1]Theoretical Division, Los Alamos National Laboratory, Los Alamos, New Mexico 87545, USA
2 Computational Science and Engineering Division, Oak Ridge National Laboratory, 1 Bethel Valley Road, Oak Ridge, TN, 37830, USA
[3] Sandia National Laboratories, Albuquerque, New Mexico 87123 USA
4 Department of Mechanical Engineering, Wayne State University, Detroit, Michigan 48202 USA
5 - School of Materials Engineering and Birck Nanotechnology Center, Purdue University, West Lafayette, Indiana, 47907 USA

SM – 1

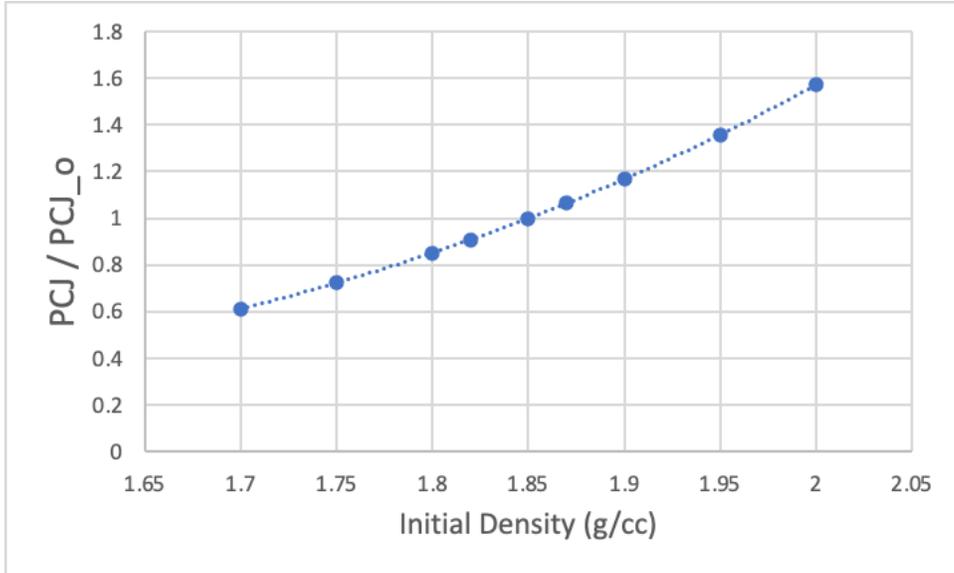

SM Figure 1: Relative change in CJ pressure for changing initial density given a single Hugoniot products curve.

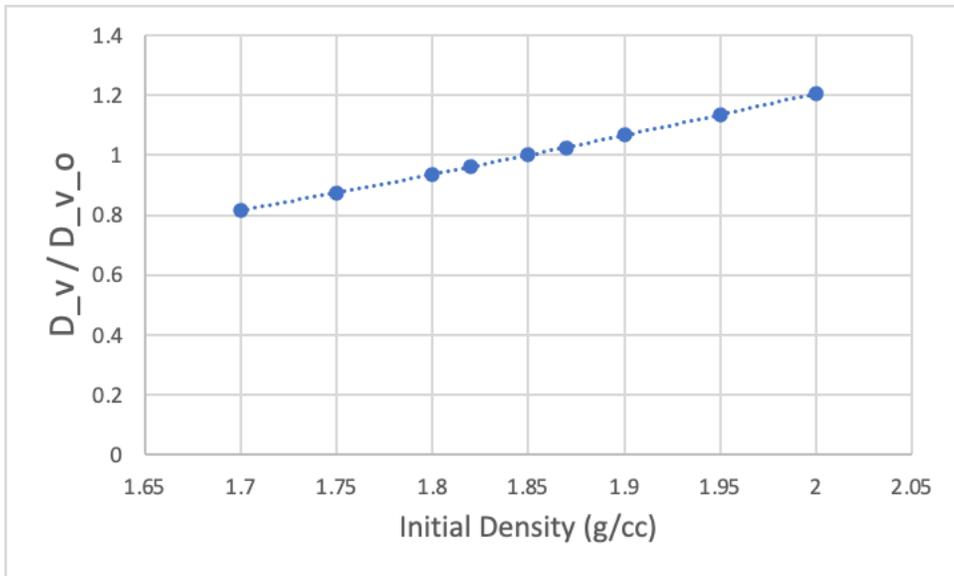

SM Figure 2: Relative change in detonation velocity for changing initial density given a single Hugoniot products curve.

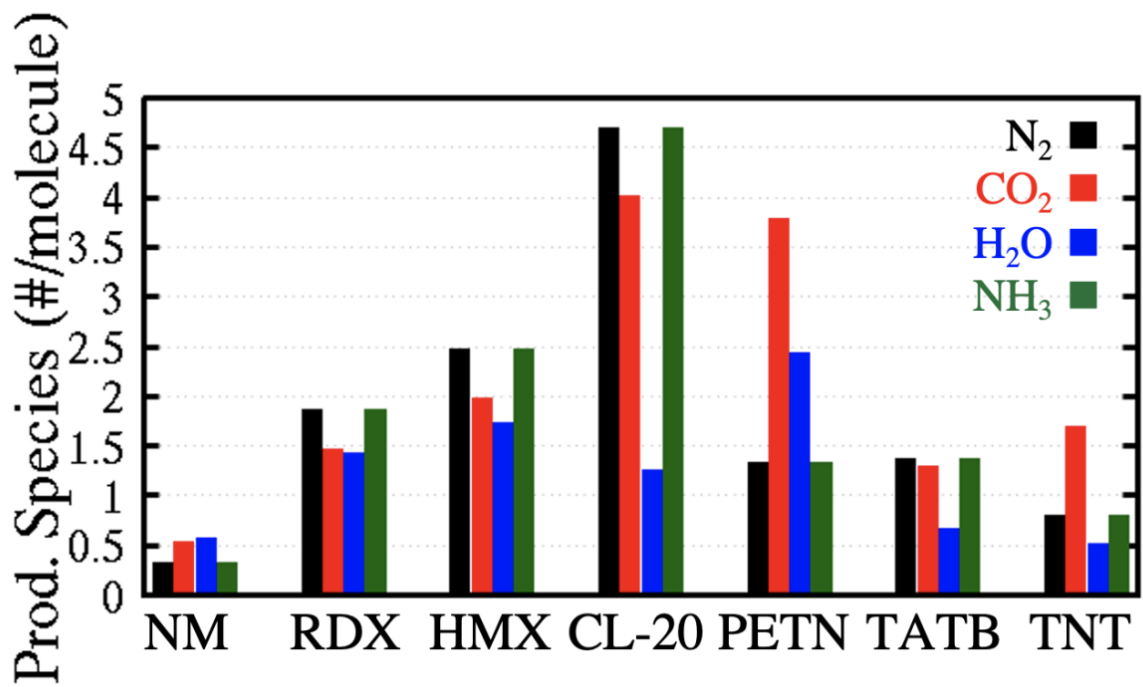

SM Figure 3: Un-scaled product species from a 2500 isothermal-isochoric decomposition reaction using NNRF Gen3.9zbl.